# Strong Chirality Suppression in 1-D correlated Weyl Semimetal $(TaSe_4)_2I$

Utkarsh Khandelwal[1+], Harshvardhan Jog[1+], Shupeng Xu[1], Yicong Chen[1], Kejian Qu[2], Chengxi Zhao[2], Eugene Mele[3], Daniel P. Shoemaker[2] and Ritesh Agarwal[1]*.

1 Department of Materials Science and Engineering, University of Pennsylvania, Philadelphia, PA 19104, USA

2 Department of Materials Science and Engineering, University of Illinois Urbana-Champaign, Champaign, IL 61820, USA

3 Department of Physics and Astronomy, University of Pennsylvania, Philadelphia, PA 19104, USA.

**The interaction of light with correlated Weyl semimetals (WSMs) provides a unique platform for exploring non-equilibrium phases and fundamental properties such as chirality. Here, we investigate the structural chirality of $(TaSe_4)_2I$, a correlated WSM, under weak optical pumping using Circular Photogalvanic Effect (CPGE) measurements and Raman spectroscopy. Surprisingly, we find that there is a loss of chirality in $(TaSe_4)_2I$ above a threshold light intensity. We suggest that the loss of chirality is due to an optically driven phase transition into an achiral structure distinct from the ground state. This structural transformation is supported by fluence-dependent Raman spectra, revealing a new peak at low pump fluences that disappears above the threshold fluence. The loss of chirality even at low optical powers suggests that the system quickly transitions into a non WSM phase, and**

**also highlights the importance of considering light-induced structural interactions in understanding the behavior of correlated systems. These studies showcase that even low excitation powers can be used to control the properties of correlated topological systems, opening up new avenues for low power optical devices.**

Over the past two decades, the exploration of quantum phenomena in topological materials has emerged as a central focus in condensed matter physics. Topological materials, characterized by nontrivial band structures[1], display novel quantum behavior including topologically protected edge states[2], topological Fermi arc surface states[3], photoinduced anomalous Hall effects[4], and nonlocal transport behavior.[5] A particular class of topological materials, Weyl Semimetals (WSMs), with broken inversion and/or time-reversal symmetry, have garnered significant attention due to their ability to host chiral massless Weyl fermions as their low energy excitations[6]. WSMs are defined by bands that linearly disperse in momentum space and traverse through nodal points known as Weyl points (WPs), which carry integer-valued topological charges, and act as sources or sinks of Berry curvature[6].

In the presence of strong electronic correlations, WSM are proposed to exhibit exotic phenomenon like topological superconductivity [7,8] or an incommensurate CDW transition[9]. Recently, theoretical studies indicate that sufficiently strong electronic correlations can gap out bulk Weyl nodes and thus break WSM states[10]. Therefore, if a system which is predicted to exhibit a WSM phase in a non-interacting single-particle picture with nonnegligible electronic correlations, it will be important to investigate the influence of electronic correlations on its predicted WSM state. $(TaSe_4)_2I$ is a 1-D WSM that offers a rare avenue to explore the effects of both electron-electron and electron-phonon interactions with topology[9]. $(TaSe_4)_2I$ was first synthesized several decades ago[11] attracting scientific interest due to its pseudogap[12] arising from the presence of polarons[13] and an incommensurate charge density wave (CDW) transition[14,15]. More recently, an observation of positive magnetoresistance in CDW phase of $(TaSe_4)_2I$ hinted towards a possible axion insulator behavior[16], but subsequent reports indicated that this effect may also be attributable to joule heating[17]. Optical excitation, which is often used to probe materials, can also influence their quantum state, especially in correlated systems even under weak excitation conditions as recently reported for excitonic insulators and CDW systems[18,19]. Light can also trigger phase transitions in materials, which suggests that the role of optical excitations in correlated topological systems such as $(TaSe_4)_2I$ need to be properly investigated. Here, we investigate how interaction with light drives $(TaSe_4)_2I$ to a fundamentally different non-equilibrium phase with restored inversion symmetry via the circular photogalvanic effect (CPGE), which is a sensitive probe of chirality. CPGE is a second order nonlinear optical response, which is expected to scale linearly with light intensity. Surprisingly, our findings reveal that in $(TaSe_4)_2I$, CPGE exhibits a strongly nonlinear behavior and saturates even under very weak optical excitation (~100 µW incident CW power), indicating the loss of chirality in the system with laser fluence in this correlated WSM. These observations, supported by fluence-dependent Raman spectroscopy, suggest that the material undergoes a light-induced structural transformation to an achiral phase (non WSM) even under very weak light excitation.

The crystal lattice structure of $(TaSe_4)_2I$ in its high-temperature phase adopts a body-centered tetragonal unit cell with dimensions a = b = 9.531 Å and c = 12.824 Å **(Fig. 1a-b)**. Chains comprising $TaSe_4$ units extend in parallel along the c-axis, revealing a distinctive alignment of Se around the Ta. This gives rise to a chiral helix structure featuring a non-crystallographic screw axis. The chains are weakly linked with iodine atoms, owing to which the crystal cleaves along the 110 planes of the lattice.

Initially studied in conventional semiconductors[20,21], CPGE is a $\chi^{[2]}$ response, which manifests as short circuit current or open circuit voltage due to the asymmetric scattering of carriers upon absorption of circularly polarized photons[20]. To the lowest order, the presence of CPGE requires broken inversion symmetry and it manifests in materials with gyrotropic point groups- $C_1$, $C_2$, $C_s$, $C_{2v}$, $C_4$, $C_{4v}$, $C_3$, $C_{3v}$, $C_6$, $C_{6v}$, $D_2$, $D_4$, $D_{2d}$, $D_3$, $D_6$, $S_4$, $T$ and $O$. [22] The observed photovoltage as a function of the angle between the quarter waveplate and the incident polarization of the laser ($\theta$) can be fitted to the following equation[19],

$$V_{tot} = V_{CPGE} \sin(2\theta) + V_{LPGE} \sin(4\theta + \delta) + V_0$$

where $V_{CPGE}$ represents the magnitude of the CPGE voltage, $V_{LPGE}$ denotes the magnitude of the Linear Photogalvanic Effect (LPGE) voltage with a phase shift $\delta$, and $V_0$ signifies the polarization-independent background voltage. Therefore, by changing the polarization of the incident laser with a quarter waveplate (QWP) and measuring the response, $V_{CPGE}$ and $V_{LPGE}$ can be obtained.

$(TaSe_4)_2I$ single-crystals with large chiral domains and uniform handedness were synthesized following the protocols in a previously published report[23]. These crystals possess a bulk iodine deficiency which contributes to the semiconductor/insulator resistivity behavior observed in the high temperature phase of $(TaSe_4)_2I$[23]. The crystals were mechanically exfoliated and transferred onto a $SiO_2/Si$ substrate using a PDMS assisted dry transfer technique. These crystals cleave into nanoribbons with lengths extending in the hundreds of microns along the c-axis, with widths varying from several hundred nanometers to a couple of micrometers. CPGE was measured at room temperature by shining light focused to a spot size of ~ 800 nm on a two-terminal device and collecting the voltage via lock-in technique (**Fig. 1c; see methods**). To ensure the elimination of anisotropic in-plane light momentum contribution to CPGE such as the photon drag effect,[24,25] light was incident normally on the sample. Furthermore, to eliminate Schottky field-dependent CPGE [26] arising from the electrodes, the light was incident away from the electrodes. **Figure 1d** shows the variation of photocurrent with the polarization of light incident normally onto a prototypical $(TaSe_4)_2I$ nanoribbon (width, ~300 nm). A discernible difference in photocurrent magnitude between left and right circular polarizations indicates the presence of CPGE.

Since $(TaSe_4)_2I$ crystallizes in the I422 space group, it meets the requisite condition for generating a CPGE signal due to its broken inversion symmetry. Further characterization of this response provides insights into light-induced dynamics of the system.[18,19] As the CPGE is a second-order nonlinear effect, it should scale linearly with laser pump intensity. This has been shown in previous measurements conducted on Si nanowire[26] and in WSM $MoTe_2$ at low temperatures[27]. However, laser intensity dependent CPGE measurements on the same nanoribbon as in **Fig. 1d**, shows an intriguing non-linear behavior (**Fig. 2a**). Initially, the response exhibits an increase, followed by a decrease, and eventually saturates to a nearly constant value with increasing optical power. The total photocurrent, which is dominated by the polarization-independent component, did not show a reversal behavior, and also did not saturate at higher pump powers (**Fig. 2b**). Since CPGE is a strictly symmetry dependent response, these observations suggest that the photovoltage is not limited by carriers in the material but rather by the modification of the material's symmetry induced by light.

The nonlinear behavior of CPGE with optical intensity observed in $(TaSe_4)_2I$ warrants further investigation. To further probe this phenomenon, we varied the excitation wavelength to study its effect on CPGE (**SI fig-1**). We did not see any significant variation with the wavelength and the response was consistent in the visible excitation range. No hysteresis of the signal was observed when the laser was cycled between high and low intensities. Notably, the observed behavior was influenced by the relative spot size of the laser compared to the width of the sample. **Fig. 3a** shows CPGE data from another nanoribbon device where the laser spot size (~800 nm) was much smaller than the width of the nanoribbon (~2 μm). Here, it was observed that the CPGE response increased linearly and then saturated with the pump fluence. In general, when the spot size exceeded the width of the nanoribbon, a distinct reversal in CPGE followed by saturation was observed with increased pump fluence **(Fig. 2a)**. However, when the spot size was smaller than the width of the nanoribbon, no reversal in CPGE was observed; instead, the signal exhibited an increase followed by saturation (**Fig. 3a).**

The absence of pump wavelength dependence on the CPGE indicates that the phenomenon is not primarily caused by the unique band structure of the material or any specific electronic transition. The saturation of the CPGE signal with pump fluence hints at an optically induced phase transition within the material, altering its chirality. To understand this relationship further, we performed simulations where a Gaussian sampled beam spot illuminated a nanoribbon, with its length much greater than the width and the spot size of the focused laser beam **(Fig. 3b)**. In our simulations, we varied the ratio of width of the nanoribbon and the spot size and implemented a criterion that light intensity above a certain threshold ($I_{Th}$) rendered the material achiral, thus not contributing to CPGE. Consequently, the CPGE signal originated solely from the less intense portion of the incident gaussian beam spot.

Results of the simulations gave further insight into the CPGE response (**Fig. 3c**) where we found a notable influence of sample geometry on the CPGE response with laser pump fluence. First, we find that the model which accounts for the loss of CPGE signals from areas of the sample with laser fluence above the threshold ($I_{Th}$) could replicate the general features of our experimental observations. Consistent with experimental findings, the model shows that when the width of the sample (w) is substantially larger than the laser beam spot size (σ) (i.e., the 2D limit) we observe only a change in slope in the CPGE signal, accompanied by saturation, as seen in **Fig. 3a**. However, a distinct hump in the CPGE signal emerges (**Fig. 2a**) when the width of the sample is much smaller or about the size of the beam spot (i.e., quasi-1D limit). In essence, these simulation results highlight the intricate interplay of sample dimensionality and experimental length scales, such as the laser spot size on the resulting CPGE response. They validate our experimental findings, emphasizing that the saturation of the CPGE signal with pump fluence reflects an optically induced chirality suppression in the material.

The idea of light induced chirality suppression is intriguing. To understand the underlying mechanisms behind the observed effect, we focus on the crystal structure of the material[28,29]. The selenium atoms form a rectangular antiprismatic arrangement around the central tantalum atom (**Fig. 4a**). These rectangles undergo a 45° rotation to minimize repulsion between occupied π and π* orbitals of selenium[29]. Due to the unique *4e* position (in Wyckoff notation) of the iodine within the crystal lattice, it bonds asymmetrically with the selenium atoms. This positioning places it closer to one selenium atom (Se(2)) than the other (Se (1)), resulting in the formation of these rectangles (**Fig. 4b**). The chirality in the $(TaSe_4)_2I$ structure arises from the rotation of these rectangles by 45° with respect to each other forming a nearly eight-fold screw of $Se_4$ rectangles.

It has been shown previously[23], that although most of the iodine electronic density remains at the *4e* Wyckoff position, a significant portion resides at the *2a* Wyckoff position[23] in the $(TaSe_4)_2I$ crystal. The *2a* position lies at the intersection of a four-fold rotation axis and several perpendicular two-fold rotation axes, situated at the corner of the body-centered tetragonal unit cell, and remains symmetric to both Se (1) and Se (2). In a correlated system, it is possible that the excess carriers generated due to photoexcitation can lead to significant redistribution of electronic density. In particular, the weakly bonded iodine atoms may experience a shift towards the more symmetric *2a* position. As iodine atoms migrate, the symmetry of the crystal lattice undergoes a significant transformation to a more symmetric transient arrangement of selenium atoms around tantalum, resulting in the formation of a square antiprismatic structure **(Fig. 4c)**. This introduces mirror planes passing through the tantalum chains, fundamentally changing the crystal's symmetry properties **(Fig. 4d)**, and the material transitions from a ground state with chiral $D_4$ into an excited state with achiral $D_{4h}$ point group.

The CPGE measurements can be explained by the chirality transition in $(TaSe_4)_2I$. As the intensity of the incident light increased beyond the threshold intensity, part of the wire that experiences $I > I_{Th}$ enters the achiral excited state and stops contributing to the CPGE signal. The signal only comes from the shoulder of the beam spot, where the $(TaSe_4)_2I$ is still in the chiral ($D_4$) phase because of reduced intensity. As seen in our simulations, this can lead to the eventual saturation of the CPGE.

To further corroborate the claim of light induced structural transition in this material, we employed Raman spectroscopy, which is also a probe of lattice structure,[30] symmetry structural changes. Numerous Raman studies have investigated the temperature-dependent[31,32] behavior of $(TaSe_4)_2I$ to elucidate the charge density wave (CDW) transition. These studies are generally carried out at higher pump intensities, and the pump fluence dependence of the Raman peaks has not been reported. Drawing insights from our CPGE measurements at low laser powers, we conducted a pump intensity-dependent Raman spectroscopy analysis of the material **(Fig. 4e; see methods)**. A peak at 271 cm$^{-1}$ is present at all fluences which corresponds to the mode associated with Se-Se bond vibration in $(TaSe_4)_2I$.[31] Notably, we observed the emergence of a new peak at 256 cm$^{-1}$ at low fluence levels, which has not been observed previously in the Raman spectrum of $(TaSe_4)_2I$. As the pump fluence was increased, this mode gradually became Raman inactive, suggesting a structural transformation in the crystal. Based on the bonding distances[28] observed in this material, a possible explanation for the 256 cm$^{-1}$ peak is the vibration of the Iodine-Selenium (I-Se) bond. As the pump intensity increases, the relaxation of the I-Se bond results in a decline in Raman intensity of this peak. Eventually, beyond a certain threshold intensity, the bond between iodine and selenium ceases to exist, leading to the disappearance of the peak. It is important to note that this new symmetry group under illumination, $D_{4h}$, possesses the same irreducible representation of Raman tensors as the previously presumed $D_4$ group[33,34]. Hence, it is plausible that previous Raman studies misidentified the crystal's symmetry as $D_4$ instead of $D_{4h}$ due to the similarities in their Raman tensors.

Our identification of an achiral nonequilibrium phase in $(TaSe_4)_2I$ under weak optical excitation represents a new understanding of this material. The suppression of chirality caused due to structural transformation of the lattice will modify the band structure. $(TaSe_4)_2I$ under optical illumination, due to restored inversion symmetry, and will not remain a Weyl semimetal. In a prior investigation using time resolved ARPES[35], a noticeable shift in bands was detected following photoexcitation using a strong pulsed laser for several hundred microseconds, which was ascribed to the accumulation of lattice temperature and limited thermal conductivity within the material. In a separate analysis based on ARPES[36], it was claimed that weak photoexcitation transiently transforms the bipolaronic ground state into polarons. Furthermore, saturation in the femtosecond laser induced THz signal[37] were linked to "exotic" electronic phases, although any evidence regarding how this change in electronic phase affects the symmetry-dependent $\chi^{[2]}$ response was lacking. However, given our results demonstrating the sharp change in the underlying symmetries of the high temperature phase of $(TaSe_4)_2I$ in the presence of weak optical excitation, it becomes imperative to reconsider these investigations. The structure and transport properties of this novel non-equilibrium phase are beyond the scope of current work and present an exciting direction for future research.

In conclusion, our study reveals the intriguing response of $(TaSe_4)_2I$ under optical excitation. We find that $(TaSe_4)_2I$ undergoes a suppression in chirality and transforms into a unique non-equilibrium achiral phase belonging to the $D_{4h}$ point group. We observe an optical intensity-dependence in the CPGE response which saturates at very low excitation laser powers. Our findings are further supported by fluence-dependent Raman spectra, revealing a new peak at low pump fluences previously unreported in $(TaSe_4)_2I$, which disappears above a threshold intensity. This observation demonstrates how optical phase control can be achieved using optical intensities as low as 10 microwatts. Our work highlights the importance of considering structural transformations induced by weak optical pumping in understanding the behavior of correlated WSMs.

## Methods

The detailed synthesis protocol is described in a previously published report[23]. $(TaSe_4)_2I$ single crystals were synthesized using chemical vapor transport. Stoichiometric amounts of Ta wire (99.9%), Se powder (99.999%) and I shot (99.99%) were loaded into a fused silica tube, which was sealed under a vacuum and heated with a source temperature of 600 °C and sink temperature of 500 °C for 10 days.

The scotch tape assisted mechanical exfoliation technique was used to cleave the crystal and transfer it onto a SiO$_2$/Si substrate with lithographically defined markers. Because of the weak inter-chain bonding in the material, the crystal cleaves as a bundle of chains, extending along the c-axis. To perform the polarization dependent photocurrent measurements, two terminal devices were fabricated using standard e-beam lithography and Ti/Au electrode were deposited with a total thickness of ~200 nm **(Fig. 1c).** A tunable supercontinuum laser (NKT Photonics with a SuperK VARIA filter) with a gaussian laser beam profile was used as an excitation source with the wavelength of 650 nm, unless otherwise mentioned and focused on the material using a home-build optical microscope. The photovoltage signal was measured with a lock-in amplifier locked at the frequency of the optical chopper. To measure the CPGE directly, we locked the lock-in amplifier at the frequency of motorized QWP rotator (frequency $\omega \sim 177$) and measured the second harmonic of the voltage recorded. This is a consequence of equation (1) of the main text.

Raman Spectroscopy was performed with a commercial NT-MDT NTEGRA setup on bulk (TaSe$_4$)$_2$I crystal using echelle grating with 75 lines/mm with spectral resolution of 0.1 cm$^{-1}$. The excitation laser of 660 nm with a peak power of 300 mW at the source was focused with 100X 0.7 N.A. objective lens on the sample. The light was incident and collected along the a-axis of the crystal. Incident power was varied by a neutral-density filter in the setup. The was optimized for different power to avoid damage to the crystal.

## Acknowledgements

Optical and transport measurements and simulations were supported by he National Science Foundation (NSF-QII-TAQS-#1936276 and NSF-2230240) and by the US Air Force Office of Scientific Research (award# FA9550-20-1-0345). This work was partially supported by the King Abdullah University of Science & Technology (OSR-2020-CRG9-4374.3). Device fabrication and characterization work was carried out in part at the Singh Center for Nanotechnology, which is supported by the NSF National Nanotechnology Coordinated Infrastructure Program under grant NNCI-1542153. Materials synthesis was supported by the Center for Quantum Sensing and Quantum Materials, an Energy Frontier Research Center funded by the U. S. Department of Energy, Office of Science, Basic Energy Sciences under Award DE-SC0021238.

# Figures

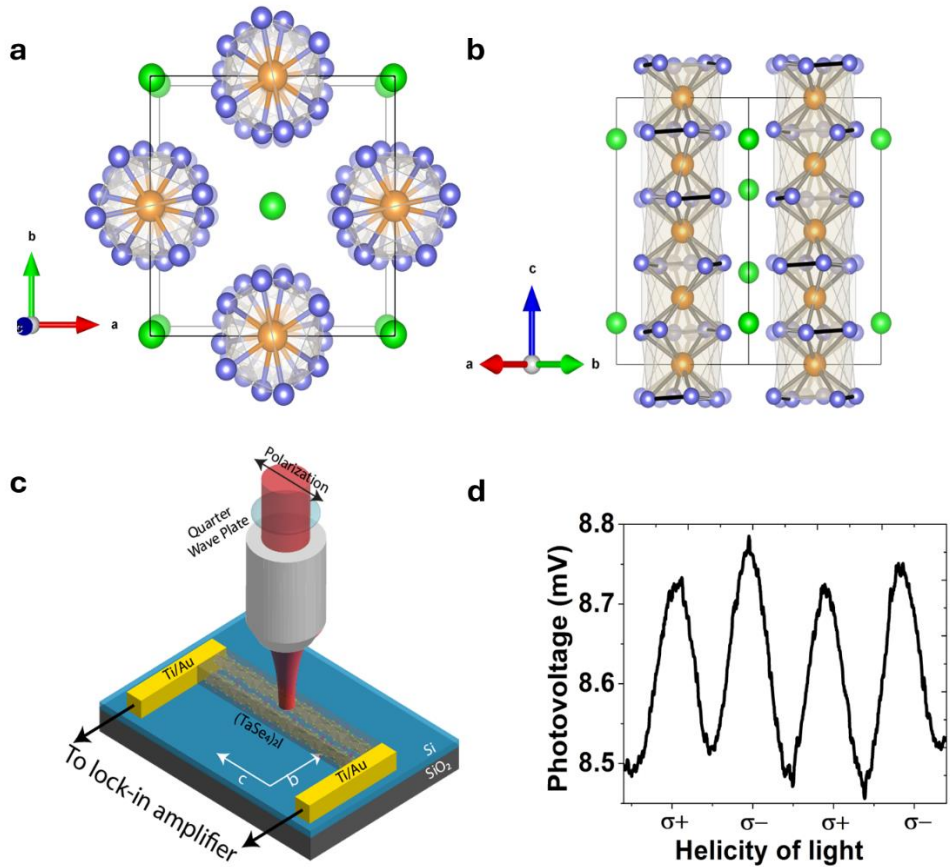

**Figure 1**: **Crystal Structure of room temperature (TaSe$_4$)$_2$I and its polarization-dependent photogalvanic response**. (a) Schematic of the crystal structure of (TaSe$_4$)$_2$I in the room temperature Weyl semimetal phase projected on the *a-b* plane. (b) Schematic of the crystal structure depicting the formation of long chains of Ta atoms along the *c*-axis (right). (c) A schematic representation of the CPGE experimental setup and the fabricated device architecture. (d) Polarization-dependent photovoltage at zero external bias observed at room temperature shows a discernible difference between the currents generated by left and right circularly polarized light, indicating the presence of CPGE.

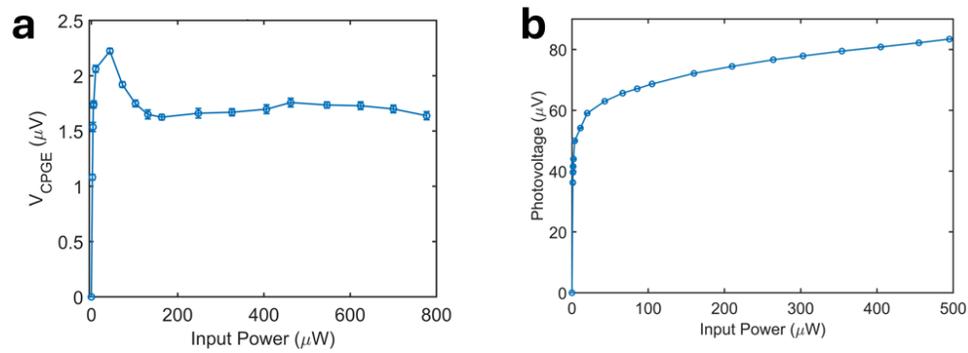

**Figure 2 : Dependence of CPGE and total photovoltage on excitation laser power for room temperature (TaSe$_4$)$_2$I.** (a) CPGE signal amplitude initially increases, reverses direction, and eventually saturates as a function of pump laser power. (b) Total photovoltage, dominated by polarization independent component, with increasing incident fluence, continues to rise with pump fluence without reaching saturation (lines are guide to the eye).

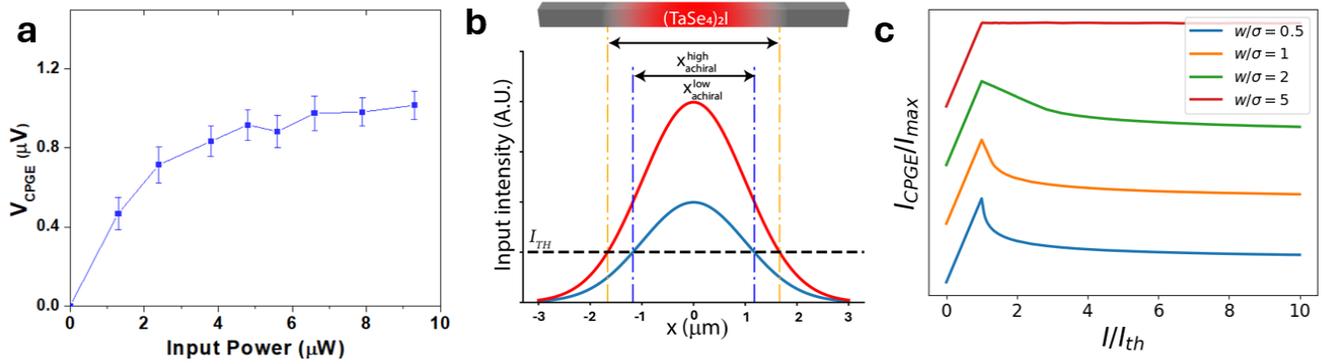

**Figure-3**: **Effect of laser spot size and sample width on the CPGE response and modelling of CPGE.** (a) Laser pump power dependent CPGE response when a gaussian laser with spot size (σ ~800 nm) is focused on a device with width (w ~2 μm); no reversal is observed. (b) Overview of the mechanism of a model for CPGE response. When part of the material ($x_{achiral}$) is exposed to laser intensity above $I_{th}$, it loses its chirality, rendering it incapable of contributing to CPGE. The CPGE only originates from the part of the sample exposed to $I<I_{Th}$ thereby ensuring a continuous CPGE. (c) The results of the simulations. It is observed that the behavior is influenced by the width of the nanoribbon. In the 2D limit, i.e., when the laser spot size is much smaller than the ribbon width (w/σ >2), no reversal is observed. Conversely, when the ribbon width is smaller or comparable to the laser spot size (Quasi-1D limit, w/σ <2), a reversal in the CPGE is observed.

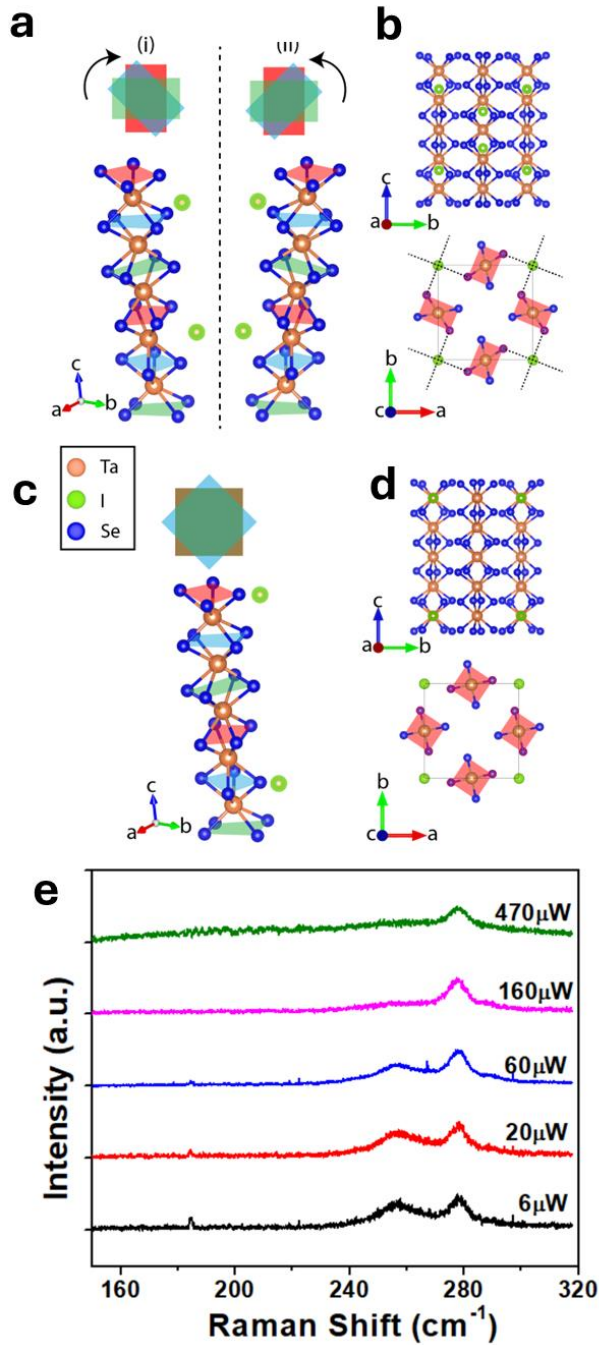

**Figure-4: Light induced chirality transformation in (TaSe$_4$)$_2$I**. (a) Room temperature crystal structure (TaSe$_4$)$_2$I showing the origin of structural chirality in (TaSe$_4$)$_2$I. The red, blue, and green planes denote the Se$_4$ rectangles which are rotated 45° with each other forming a chiral nearly eight-fold non-crystallographic screw axis (b) The b-c plane view (top) of the crystal structure and the a-b plane view (bottom) of the crystal at z =0.125 showing the preferential interaction of Iodine (4e Wyckoff position) with only with Se(2). For clarity, Se(1) is denoted in blue and Se(2) in red. (c) Schematic of the proposed crystal structure under optical illumination. Iodine atoms in the material moves to the 2a Wyckoff position, forming a square antiprismatic structure. Due to the formation of the $Se_4$ squares (red, green, and blue planes), the chirality is lost in the structure. (d) The a-b plane view of the proposed optically excited structure showing the disruption of I-Se(2) bond, leading to a symmetric arrangement of Se around the Ta and disappearance of chirality. (e) Laser power-dependent normalized Raman spectra. The mode at 271 $cm^{-1}$, corresponding to the Se-Se bond appears for all laser powers. However, the peak at 256 $cm^{-1}$ emerges at only at low laser powers but disappears beyond a certain threshold power (~60 µW), indicating a structural transformation in the material.